\documentstyle[twoside,fleqn,npb,epsfig,amsmath,here]{article}

\newcommand{\nn}{\nonumber}

\def\fmfL(#1,#2,#3)#4{\put(#1,#2){\makebox(0,0)[#3]{#4}}}

\def\thefigure{}

\title{FIRCLA, one-loop correction to $e^+e^- \rightarrow \nu \overline{\nu} H$
and basis of Feynman integrals in higher dimensions}

\author{F.~Jegerlehner~\address{Deutsches Elektronen-Synchrotron,
DESY Zeuthen, Platanenallee 6, D-15738 Zeuthen, Germany }
\addtocounter{address}{-1}
     and   O.~V.~Tarasov~\addressmark\thanks{Supported in part by DFG 
under the project FL 241/4-2} }

\allowdisplaybreaks

\begin{document}
\onecolumn{
\renewcommand{\thefootnote}{\fnsymbol{footnote}}
\setlength{\baselineskip}{0.52cm}
\thispagestyle{empty}
\begin{flushleft}
DESY 02--210 \\
November 2002\\
\end{flushleft}

\setcounter{page}{0}

\vspace*{3cm}

\begin{center}
{\LARGE\bf FIRCLA, one-loop correction to $e^+e^- \rightarrow \nu
\overline{\nu} H$ \\[3mm] and basis of} \\  
\vspace{3mm}
{\LARGE\bf  Feynman integrals in higher dimensions\footnote[3]{\noindent 
        To appear in the Proceedings of ``RADCOR 2002 - Loops and Legs 2002'', 
        Kloster Banz, 2002,
        Nucl. Phys. B (Proc.Suppl.)
}}\\
\vspace*{2cm}
Fred Jegerlehner and
Oleg Tarasov\footnote{Supported in part by DFG 
under the project FL 241/4-2}\vspace{0.5cm}\\

\vspace*{1cm}

\noindent
{\it
Deutsches Elektronen-Synchrotron DESY, Platanenallee 6, D-15738 Zeuthen,
Germany }
\end{center}

\vspace{5em}
\large
\vspace*{\fill}}

\newpage 

\begin{abstract}
An approach for an effective computer evaluation of one-loop multi-leg
diagrams is proposed. It's main feature is the combined use of several
systems - DIANA, FORM and
MAPLE.  As an application we consider the one-loop
correction to Higgs production in $ e^+ \, e^- \to \nu \,
\overline{\nu}\, H$, which is important for future $e^+ e^-$
colliders. To improve the stability of numerical evaluations a
non-standard basis of integrals is introduced by transforming
integrals to higher dimensions.\\[-6mm]
\end{abstract}

\maketitle

\section{INTRODUCTION}
Electroweak SM calculations of one-loop corrections to processes with
five and more external legs are quite demanding. The number of
diagrams to be evaluated typically is in the several thousands,
integrands of the individual diagrams are in general very lengthy and
in spite of large cancellations which take place the final result
usually comes out to be not as compact as required for an efficient and
reliable numerical evaluation. At the present time there is no
computer algebra system which allows one to perform in an efficient
way complete calculations of multi-leg one-loop corrections. For
example, FORM is very efficient in doing Lorentz algebra
but it can not handle ratios of Gram determinants depending on several
momenta and masses. To overcome this kind of problems we use a
combination of several systems and exploit their most advantageous
features. We are testing the effectiveness of our strategy in the
evaluation of the one-loop correction to the process $ e^+ \, e^- \to
\nu \, \overline{\nu}\, H$.  In the last part of this note we also will
discuss improvements possible by utilizing non-standard sets of
master-integrals.

\section{FIRCLA}
To achieve a better performance we use a combination of several tools
like DIANA ~\cite{Tentyukov:1999is}  based on C and program libraries
based on FORM and MAPLE which also act as interfaces between them.
Since the main tool of
evaluation is based on recurrence relations we called this collection
of packages FIRCLA which stands for {\it {\bf F}eynman {\bf
I}ntegral {\bf R}ecursive {\bf C}alcu{\bf LA}tor}.

FIRCLA works as follows. After the process is specified to DIANA it
invokes QGRAF~\cite{Nogueira:1991ex}, to generate all diagrams, then
constructs input expressions suitable for use by FORM and provides
additional information like types and masses of particles as well as
the relations between kinematical variables. The output from DIANA is
stored in a file which is used as an input by our FORM
package. Additionally, DIANA may be utilized to produce pictures of
all or particular diagrams as a Postscript file.

The FORM package performs the Lorentz algebra, takes traces of the
Dirac $\gamma$-matrices, transforms products of spinors times $\gamma$
matrices to a chosen basis of amplitudes, and utilizing the
algorithm~\cite{Tarasov:1996br}, reduces tensor integrals to
a combination of scalar integrals, some of them with shifted
space-time dimension. For each diagram the FORM package creates a file
with expression for further processing by our MAPLE package. The
expressions are written in MAPLE format. Each of these files also
provides information about scalar products of external momenta and the
masses of the particles.

 The values of all scalar invariants are calculated in the FORM
 package.  Relations between the momenta carried by the lines of the
 diagrams ($p_i$) and the external momenta ($q_j$) are evaluated by
 DIANA after generating a diagram and then transfered via FORM to the
 MAPLE package. In this way all useful information from DIANA can be
 transfered to the MAPLE package. The relations between momenta are
 needed, for example, to find which integrals are independent or
 equivalent. In the FORM output files, after the kinematic information
 an expression in MAPLE format follows.  It has the form of a
 sum of integrals multiplied by polynomials of scalar products, masses
 and spinor amplitudes. Integrals are just the names of MAPLE
 procedures like:\\[-3mm]
\begin{equation}
{\rm
npoint( \prod_{j=1}^N P(j,m_j^2)^{\nu_j},p,data)} \vspace*{-3mm}
\end{equation}
where $P(j,m_j^2)=(k_1-p_j)^2-m_j^2 + i \epsilon $ is an inverse
scalar propagator, $\nu_j$ its power, $N$ the number of different
propagators, the parameter $p$ specifies the shift of the space-time
dimension $D=d+2p$ and $data$ is a set of substitutions for the
momenta and scalar invariants.

In the MAPLE package integrals are evaluated separately by using the
recurrence relations algorithm for the evaluation of one-loop
integrals described in~\cite{Tarasov:1996br,fjt00} (see also~\cite{Bern:1993kr}).

If the Gram determinants of an integral are different from zero then a
set of three relations can be used: the relation for removing dots
(a dot represents one power of momentum attached to the line)
from lines reads\\[-5mm]
\begin{eqnarray}
&&2 \Delta_n \nu_j {\bf j^+} I_n^{(d)}=
  \sum^{n}_{k=1} (1+\delta_{jk})   \left( \frac{\partial \Delta_n}
 {\partial Y_{jk}} \right)
\nonumber \\
&&
~~~~\times \left[ d - \sum_{i=1}^{n} \nu_i( {\bf k^-} {\bf i^+}+1)
             \right] I_n^{(d)}.
\label{recurseJ}
\end{eqnarray}
Another relation reduces the shift of the space-time dimension and the
index of the $j$-th line simultaneously
\begin{eqnarray}
&&G_{n-1} \nu_j{\bf j^+} I^{(d+2)}_n= \nonumber \\
&& \left[ (\partial_j  \Delta_n) +\sum_{k=1}^{n}
 (\partial_j \partial_k \Delta_n)
 {\bf k^-} \right] I^{(d)}_n\;.
\label{reduceJandDtod}
\end{eqnarray}
Integrals with shifted  space-time dimension can be expressed
in terms of integrals in generic dimension by applying the formula:
\begin{eqnarray}
&&(d-\sum_{i=1}^{n}\nu_i+1)G_{n-1}I^{(d+2)}_n= \nonumber \\
&&  \left[2 \Delta_n+\sum_{k=1}^n (\partial_k \Delta_n) {\bf k^-}
  \right]I^{(d)}_n\;.
\label{reduceDtod}
\end{eqnarray}
In the above formulae the shift operators ${\bf j^{\pm }}$ etc. shift
the indices $\nu_j \to \nu_{j } \pm 1$,
\begin{eqnarray}
\Delta_n=  \left|
\begin{array}{cccc}
Y_{11}  & Y_{12}  &\ldots & Y_{1n} \\
Y_{12}  & Y_{22}  &\ldots & Y_{2n} \\
\vdots  & \vdots  &\ddots & \vdots \\
Y_{1n}  & Y_{2n}  &\ldots & Y_{nn} \nn
\end{array}
         \right|, \\
~Y_{ij}=-(p_i-p_j)^2+m_i^2+m_j^2,\nn
\end{eqnarray}
$p_j$ are combinations of external momenta  and $m_j$ is the mass of the $j$-th line ,
$\partial_j \equiv \partial / \partial m_j^2$ and
\begin{equation}
G_{n-1}\!=\! -2^n \left| \!
\begin{array}{cccc}
  p_1p_1  & p_1p_2  &\ldots & p_1p_{n-1} \\
  p_1p_2  & p_2p_2  &\ldots & p_2p_{n-1} \\
  \vdots  & \vdots  &\ddots & \vdots \\
  p_1p_{n-1}  & p_2p_{n-1}  &\ldots & p_{n-1}p_{n-1}
\end{array} \!
\right|.\nn
\label{Gn}
\end{equation}
When one of the determinants $G_{n-1}$ and $\Delta_n$ is equal to zero
it is possible to express integrals with
$n$ lines as a combination of integrals with $n-1$ lines. If $G_{n-1}=0$ then
the relation
\begin{equation}
I^{(d)}_n= -\sum_{k=1}^{n}\frac{ (\partial_j \partial_k \Delta_n)}
  {(\partial_j  \Delta_n)}~ {\bf k^-}  I^{(d)}_n
\label{Gnzero}
\end{equation}
should be applied until one of the lines of the integral
will be removed. For the integral with $n-1$ lines the Gram determinant
may be different from zero and therefore the relations
(\ref{recurseJ}), (\ref{reduceJandDtod}) and (\ref{reduceDtod}) can be used
if needed.

If $\Delta_n=0$ one should apply the relation
$$
(d-\sum_{i=1}^{n}\nu_i-1)G_{n-1}I^{(d)}_n=
  \sum_{k=1}^n (\partial_k \Delta_n) {\bf k^-} I^{(d-2)}_n,
$$
until a line of the integral will be removed. As in the previous case
if the Gram determinants for integrals with $n-1$ lines are different
from zero one can use relations (\ref{recurseJ}),
(\ref{reduceJandDtod}) and (\ref{reduceDtod}). Since the space-time
dimension of $n-1$-point integrals will be decreased by this reduction
one should subsequently apply the relation
\begin{equation}
I^{(d)}_n ~=~ -\sum_{j=1}^n \nu_j {\bf j^+}I^{(d+2)}_n ~\; ,
\label{extra}
\end{equation}
in order to increase the dimension back to $d=4-2\varepsilon$.

Our MAPLE package reads the FORM output diagram by diagram, evaluates
each integral from the input expression, adds results for the
integrals and simplifies the sum. For each diagram the result is
written in a separate file. The expression stored for each diagram is
a combination of master-integrals and spinor-amplitudes multiplied by
ratios of polynomials in scalar products of external momenta, masses
and the dimension $d$. Summing up the results from
all diagrams is performed by a separate program.


\section{\protect $O(\alpha)$~CORRECTION~TO~$e^+ e^- \rightarrow
\overline{\nu} \nu H$}
In order to check the effectiveness of our method of evaluating
Feynman diagrams we calculated the one-loop correction to Higgs
production via\\[-3mm]
$$e^+(q_1) e^-(q_2) \rightarrow \overline{\nu}(q_3) \nu (q_5) H(q_4)\;.$$  
This process will be important at future $e^+e^-$ colliders. In the
tree approximation it was considered in~\cite{eennh}. At one-loop
order recently in~\cite{Belanger:2002me}.

At $e^+e^-$ linear colliders operating in the  300--800 GeV
energy range,  the main
production mechanisms for SM-like Higgs particles are
\begin{eqnarray}
e^{+}e^{-} & \rightarrow & (Z) \rightarrow Z+H \nonumber \\
e^{+}e^{-} & \rightarrow & \bar{\nu}
\ \nu \ (WW) \rightarrow \bar{\nu} \ \nu \ + H \nonumber \\
e^+ e^- & \rightarrow & e^+ e^- (ZZ) \rightarrow
e^+ e^- + H \nonumber \\
e^+ e^- & \rightarrow & (\gamma,Z) \rightarrow t\bar{t}+H
\nonumber
\end{eqnarray}
i.e., ``Higgs-strahlung'', $WW$-fusion (see figure),
$ZZ$-fusion, and ``radiation~off-top'', respectively.
The first two processes\\
\unitlength 1mm
\begin{picture}(40,20)
  \put(-0.3,-7){\includegraphics{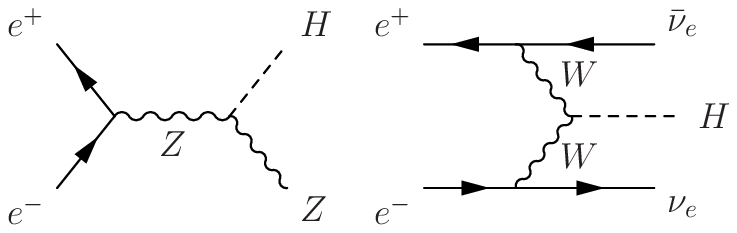}}
\end{picture}

\vspace{7mm} \noindent
are the dominant ones, in particular at energies above 500 GeV.\\
The Higgs-strahlung cross section for large $s$ goes like $\sim 1/s$
and dominates at low energies. In contrast, the $WW$-fusion mechanism
exhibits a cross section growing like $\sim \log(s/M_H^2)$ and
therefore dominates at high energies. At $\sqrt{s}
\sim 500$ GeV, the two processes have approximately the same
cross sections.  The relevance of one-loop corrections was considered
in~\cite{Kniehl:2001jy}.

In order to generate all required diagrams a 'technical SM model' with
two leptons doublets (in order to distinguish external leptons from
leptons in loops) and one quarks doublet was considered. The
contributions of the missing fermions have been obtained by adding up
the corresponding results with the appropriate masses.

We put $m_e=0$ but keep all other masses
$m_{\mu},~~m_b~~m_W,~~m_Z,~~m_t,~~m_H $ different from zero. The
results for the pentagon diagrams depend on 5 scalar invariants. All
calculations were done for arbitrary values of the gauge parameters
$\xi_W,~\xi_Z,~\xi_{\gamma}$.  In the specified model 326 diagrams
contribute to the one-loop correction. Out of these 15 are
pentagon type diagrams.

The results have been stored for each diagram separately in a file. Its
size is $\sim 100 ~MB$ ( $\sim 5 MB $ in MAPLE format). The
expression for each diagram has the form $$ D_k = \sum_{i,j} A_{i,j}
I_i O_j\; , $$ where $I_i$ are master integrals and $O_j$ are 24 spinor
amplitudes.

After summing all diagrams we find that the result is gauge invariant
as it must be. Each pentagon diagram separately gives a gauge invariant
scalar 5-point integral - 15 in total. We observed a huge reduction in
the number of scalar n-point (n=4,3,2) integrals $D_0$,
$C_0$,$B_0$. For example, from 240 integrals $C_0$ only 79 remain in
the final answer.  It should be mentioned that at the initial stage
before taking into account the symmetries of the integrals the total
amount of master integrals $C_0$ with different values of momenta and
masses was more than 500.  A special routine has been added to the
MAPLE package which recognizes the symmetries of integrals. By
permuting masses and momenta into lexicographical order, taking
into account momentum conservation, one obtains a set of independent
standard integrals and hence a much more compact representation.

The final gauge invariant result is stored in a file of about 1.2 MB
in MAPLE format. Thus, it is still rather large for further numerical
evaluations. Working out a more compact representation of the result,
mainly by optimizing ratios of huge polynomials depending on several
variables is one of the problems.

To obtain the complete radiative correction to the observable
differential cross section we still have to add the bremstrahlung
correction to our result. This work is in progress.

One observation we made is that in the final result the pentagon
diagrams yield terms of the form
\begin{equation}
\frac{1}{(d-4)} I_{11111}^{(d)}\; .
\end{equation}
In fact, second rank $n$-point tensor integrals produce terms of the
form $\frac{1}{(d-n+1)}I^{(d)}_{11...1}$ with a coefficient which is
singular at $d=4$ for $n=5$. In our case, these terms originate from
2nd rank tensor integrals like
\begin{equation}
\int d^d k_1 \frac{
(\overline{U} {\ldots} \hat{k_1}{\ldots} U)
(\overline{V}{\ldots}  \hat{k_1}{\ldots} V)}
{c_1 c_2 c_3 c_4 c_5}\;.
\label{spinors}
\end{equation}
The tensor integral here can be written in terms of scalar
integrals with shifted dimension:
\begin{eqnarray}
&&\int d^dk_1 \frac{k_{1\mu}k_{1\nu}}{c_1c_2c_3c_4c_5} =
 - \frac12  g_{\mu \nu} I_{11111}^{(d+2)}
\nonumber \\
&&
   + 2 p_{1 \mu} p_{1\nu} I_{31111}^{(d+4)}
 + 2 I_{13111}^{(d+4)} p_{2\mu} p_{2 \nu}
\nonumber \\
&&
       +  2 I_{11311}^{(d+4)} p_{3\mu} p_{3 \nu}
       +  2 I_{11131}^{(d+4)} p_{4\mu} p_{4 \nu}
\nonumber\\
&&
       +  I_{22111}^{(d+4)} \{ p_1,p_2\}_{\mu \nu}
       +  I_{21211}^{(d+4)}\{ p_1,p_3\}_{\mu \nu}
\nonumber \\
&&
+ I_{21121}^{(d+4)} \{ p_1,p_4\}_{\mu \nu}
       +I_{12211}^{(d+4)}\{ p_2,p_3\}_{\mu \nu}
\nonumber \\
&&
+  I_{12121}^{(d+4)}  \{ p_2,p_4\}_{\mu \nu}
       +  I_{11221}^{(d+4)}  \{ p_3,p_4\}_{\mu \nu} \nonumber
\end{eqnarray}
where $
\{ p_1,p_2\}_{\mu \nu}= p_{1\mu} p_{2 \nu} + p_{1 \nu} p_{2\mu}$.

Different tensor structures give contributions to different of the
abovementioned spinor amplitudes.  By using the recurrence relations
one gets
$$ I^{(d+2)}_{11111}=\frac{2\Delta_5}{(d-4) G_4}I^{(d)}_{11111}
+\sum_{k=1}^{5}\frac{\partial_k \Delta_5}{(d-4) G_4} {\bf k^-} I_5^{(d)}\;.
$$
The integral at the left hand side is UV and IR finite, however, in front of
the integrals at the right hand side the spurious pole $1/\varepsilon$
appears. So it looks like we have to evaluate the $\varepsilon$ term
in the expansion of the pentagon integrals. There are no such
problems for the 3- and 4-point functions.

In fact multiplying (\ref{spinors}) by the Born term, summing over
polarizations and taking the traces removes the problematic $ 1/(d-4)$
terms.  However, if we would attempt to calculate the amplitude
numerically before squaring it, we would have to work out first it's
$\varepsilon$-expansion. A possibility to avoid this problem is
to use a different basis of master-integrals in higher dimensions.

\section{MASTER-INTEGRALS IN HIGHER DIMENSIONS}
The idea to express one-loop tensor integrals in terms of integrals
with shifted dimension was proposed in~\cite{Davydychev:1991va} and
later in~\cite{Bern:1993kr}. Recurrence relations which allow us to
reduce any one-loop integral with shifted dimension to a combination
of integrals in generic dimension were given in~\cite{Tarasov:1996br}
and~\cite{fjt00}. These were extended to massless integrals with
both Gram determinants zero in~\cite{Binoth:1999sp}.

In~\cite{Campbell:1996zw} it was discovered that one-loop integrals in higher
dimensions provide better numerical stability.

A simple formula expressing any $n$-point integral
in terms of integrals in higher dimensions was given
in~\cite{fjt00}. It reads
$$
I^{(d)}_n =\frac{ (d-n+1)G_{n-1}}{2 \Delta_n } I^{(d+2)}_n
-\sum_{k=1}^n \frac{(\partial_k \Delta_n)}{2 \Delta_n } {\bf
k^-}I^{(d)}_n\;.
$$
The improved stability of numerical integrations of integrals in
higher dimensions can be seen from their integral representation
$$
I_{n}^{(d)} = \Gamma \left(n-\frac{d}{2}\right)
\int_0^1  dx_1 {\ldots} \int_0^1 d x_{n-1}~J_n h_n^{(  d/2-n )},
$$
where
\begin{eqnarray}
J_2&=&1\;,~\cdots ,~~J_n = x_{n-2}x_{n-3}^2 {\ldots} x_1^{n-2}\;,
\nonumber \\
h_2 &=&p_{12}^2 x_1^2+m_2^2 - x_1 (p_{12}^2-m_1^2+m_2^2)\;,
\nonumber \\
h_3 &=& -x_1  x_2  (1-x_1)  p_{13}^2 - x_1^2  x_2  (1-x_2)  p_{12}^2
\nonumber \\
&&- x_1  (1-x_1)  (1-x_2)  p_{23}^2+x_1  x_2  m_1^2
\nonumber \\
&& + x_1  (1-x_2)  m_2^2+(1-x_1)  m_3^2\;.\nonumber
\end{eqnarray}
If we transform integrals to the dimension  $D=d+2n-2$
(assuming that $d=4-2\varepsilon$) then the expansion of $I_n^{(d+2n-2)}$
for small $\varepsilon =2-d/2$ reads
$$I_n^{(d+2n-2)}=\Gamma\left(1+\varepsilon\right)
\left[-\frac{s_n}{\varepsilon} -s_n - R^{1}_n + O(\varepsilon) \right],
$$
where\\[-3mm]
\begin{equation}
s_n=\int_0^1{\ldots}  \int_0^1 \{dx\}~J_n h_n
= \frac{-\sum p_{ij}^2}{(n+1)!}  + \frac{1}{n!} \sum_{j=1}^n m_j^2,
\nonumber \vspace*{-2mm}
\end{equation}
\begin{equation}
R^{1}_n=\int_0^1{\ldots}  \int_0^1 \{dx\}~J_n  h_n \ln h_n. \nonumber
\end{equation}

\vspace{2mm} \noindent
If we transform the integrals to dimension $D=d+2n-4$  then
the expansion of $I_n^{(d+2n-4)}$ at small $\varepsilon$ reads
$$I_n^{(d+2n-4)}=\Gamma\left(1+\varepsilon\right)\left[
-\frac{1 + \varepsilon}{(n-1)! \varepsilon} - R^{0}_n + O(\varepsilon)
\right],
$$ where\\[-3mm] 
$$ R_n^0=\int_0^1{\ldots} \int_0^1 \{dx\}~J_n \ln h_n.  $$ 
We see that in the integrals $R_n^{0,1}$ there are no polynomials 
in the denominator and this is why these integrals in higher dimensions
are more suitable for a direct numerical evaluation.

As an example we give the relations for some integrals in the
logarithmic basis.  In the latter for $n>2$ we express the integrals
$I^{(d)}_n$ in terms of integrals $I^{(d+2n-4)}_n$:
\begin{eqnarray}
&&2\lambda_{123}I_3^{(d)} =(d-2)g_{123}I_3^{(d+2)}
- \partial_1 \lambda_{123}I_2^{(d)}(23) \nonumber \\
&&
    - \partial_2 \lambda_{123}I_2^{(d)}(13)
    - \partial_3 \lambda_{123}I_2^{(d)}(12),\nn
\end{eqnarray}
\begin{eqnarray}
&&4 \lambda_{1234}^2 \lambda_{123}\lambda_{124}\lambda_{134}
                   \lambda_{234}I_4^{(d)} =
(d-1)(d-3)
\nonumber \\
&&
\times g_{123}^2 \lambda_{123}\lambda_{124}\lambda_{134}
                   \lambda_{234}I_4^{(d+4)}   \nonumber \\
&&+f^{(3)}_{1234} + f^{(3)}_{2134} + f^{(3)}_{3124} + f^{(3)}_{4123}
 + f^{(2)}_{1234}
\nonumber \\
&&    + f^{(2)}_{3124} + f^{(2)}_{4123}
      + f^{(2)}_{3214} + f^{(2)}_{4213} + f^{(2)}_{4312} \nn
\end{eqnarray}
where
\begin{eqnarray}
&& f^{(2)}_{i_1i_2i_3i_4} =
( \partial_{i_1} \lambda_{i_1i_2i_3i_4}
   \partial_{i_2} \lambda_{i_2i_3i_4}
   \lambda_{i_1i_3i_4}
\nonumber \\
&&~~~ + \partial_{i_2} \lambda_{i_1i_2i_3i_4}
   \partial_{i_1} \lambda_{i_1i_3i_4}
   \lambda_{i_2i_3i_4})
\nonumber \\
&&~~~
\times   \lambda_{i_1i_2i_3i_4} \lambda_{i_1i_2i_3}
   \lambda_{i_1i_2i_4}   I_2^{(d)}(i_3i_4)
\nonumber \\
&&
 f^{(3)}_{i_1i_2i_3i_4}=
   - \partial_{i_1} \lambda_{i_1i_2i_3i_4}
    ((d-2)   g_{i_2i_3i_4} \lambda_{i_1i_2i_3i_4}
\nonumber \\
&&~~~
  +(d-3)   g_{i_1i_2i_3i_4}   \lambda_{i_2i_3i_4})
\nonumber \\
&&~~~ \times
 \lambda_{i_1i_2i_3} \lambda_{i_1i_2i_4}
   \lambda_{i_1i_3i_4}  I_3^{(d+2)}(i_2i_3i_4).\nn
\end{eqnarray}

where the following notation is used:
\begin{equation}
\lambda_{i_1 ... i_n}= \Delta_n,~~~~~~g_{i_1...i_n}= G_{n-1}\; . \nn
\end{equation}
In these expression we see multiple occurrence of Gram determinants
and their derivatives. This fact may lead for some problems for
kinematical regions where Gram determinants are close to zero.
However outside these regions we expect good numerical stability of
these integrals. In fact there is a correspondence between
the integrals in higher dimensions and the method advocated
in~\cite{Passarino:2001jd} for a direct numerical calculation of loop
integrals. For the one-loop case, (\ref{reduceJandDtod})
and (\ref{reduceDtod}) are the explicite solutions of the relations
which are derived in~\cite{Passarino:2001jd} by exploiting the
Bernstein-Tkachov theorem.

\end{document}